\documentclass[a4paper,conference]{IEEEtran}

\IEEEoverridecommandlockouts

\usepackage{graphicx,epsfig,amsmath,amsbsy,amssymb,citesort,verbatim,url,algorithm,algpseudocode,soul}
\usepackage[usenames]{color}
\usepackage{setspace,amsfonts,epsf,url}
\usepackage{subfigure}
\usepackage{bbm}
\usepackage[normalem]{ulem}

\newcommand{\x}{{\bf x}}
\newcommand{\A}{{\bf A}}
\newcommand{\y}{{\bf y}}
\newcommand{\z}{{\bf z}}
\newcommand{\f}{{\bf f}}

\newtheorem{mydef}{Definition}
\newtheorem{mylemma}{Lemma}
\newtheorem{corollary}{Corollary}

\newtheorem{CONJ}{Conjecture}
\begin{document}

\title{Performance Trade-Offs in Multi-Processor\\Approximate Message Passing}

% conference papers do not typically use \thanks and this command
% is locked out in conference mode. If really needed, such as for
% the acknowledgment of grants, issue a \IEEEoverridecommandlockouts
% after \documentclass

% for over three affiliations, or if they all won't fit within the width
% of the page, use this alternative format:
%
\author{\IEEEauthorblockN{Junan Zhu,\IEEEauthorrefmark{1}
Ahmad Beirami,\IEEEauthorrefmark{2} and
Dror Baron\IEEEauthorrefmark{1}}
\IEEEauthorblockA{\IEEEauthorrefmark{1}Department of Electrical and Computer Engineering, North Carolina State University, Email: \{jzhu9,barondror\}@ncsu.edu}
\IEEEauthorblockA{\IEEEauthorrefmark{2}Research Laboratory of Electronics, Massachusetts Institute of Technology, Email: beirami@mit.edu}
\thanks{The work was supported by the
National Science Foundation under the Grant CCF-1217749.}
}

% make the title area
\maketitle

% As a general rule, do not put math, special symbols or citations
% in the abstract
\begin{abstract}
We consider large-scale linear inverse problems in Bayesian settings.
Our general approach follows a recent line of work that applies the
approximate message passing (AMP) framework in multi-processor (MP)
computational systems by storing and processing
a subset of rows of the measurement
matrix along with corresponding measurements at each MP node.
In each MP-AMP iteration, nodes of the MP system and its fusion center
exchange lossily compressed messages pertaining to their estimates of the input.
There is a trade-off between the physical costs of the reconstruction process including computation time, communication loads, and the reconstruction quality, and it is impossible to simultaneously minimize all the costs. 
We pose this minimization as a multi-objective optimization problem (MOP), and study the properties of the best trade-offs (Pareto optimality) in this MOP. We prove that the achievable region of this MOP is convex, and conjecture how the combined cost of computation and communication scales with the desired mean squared error. These properties are verified numerically.
\end{abstract}

% no keywords

% For peer review papers, you can put extra information on the cover
% page as needed:
% \ifCLASSOPTIONpeerreview
% \begin{center} \bfseries EDICS Category: 3-BBND \end{center}
% \fi
%
% For peerreview papers, this IEEEtran command inserts a page break and
% creates the second title. It will be ignored for other modes.
\IEEEpeerreviewmaketitle

\section{Introduction}
%--------------------
Many scientific and engineering problems~\cite{DonohoCS,CandesRUP} can be approximated
as linear systems of the form
\begin{equation}
\y = \A\x + \z,
\label{eq:matrix_channel}
\end{equation}
where $\x\in\mathbb{R}^N$ is the unknown input signal, $\A\in\mathbb{R}^{M\times N}$ is the matrix that characterizes the linear system, and $\z\in\mathbb{R}^M$ is measurement noise. The goal is to estimate $\x$ from the noisy measurements $\y$ given $\A$ and statistical information about $\z$. Alternately, one could view the estimation of $\x$ as fitting or learning a linear model for the data comprised of $\y$ and $\A$.

When $M\ll N$, the setup~\eqref{eq:matrix_channel} is known as compressed sensing (CS)~\cite{DonohoCS,CandesRUP}; by posing a sparsity or compressibility
requirement on the signal,
it is indeed possible to accurately recover $\x$ from the ill-posed linear system~\cite{DonohoCS,CandesRUP} when the number of measurements $M$ is large enough, and the noise level is modest. However, we might need $M>N$ when the signal is dense or the noise is substantial.

Approximate message passing (AMP)~\cite{DMM2009,Montanari2012,Bayati2011} is an iterative framework that solves  linear inverse problems by successively decoupling~\cite{Tanaka2002,GuoVerdu2005,GuoWang2008} matrix channel problems into scalar channel denoising
problems with additive white Gaussian noise (AWGN). AMP has received considerable attention, because of its fast convergence and the state evolution (SE) formalism~\cite{DMM2009,Bayati2011}, which offers a precise
characterization of the AWGN denoising problem in each iteration.
In the Bayesian setting, AMP often achieves the minimum mean squared error (MMSE)~\cite{ZhuBaronCISS2013,Krzakala2012probabilistic} in the limit of large linear systems.

In real-world applications, a multi-processor (MP) version of CS could be of interest, due to either storage limitations in each individual processor node, or the need for fast computation. This paper considers  multi-processor CS (MP-CS)~\cite{Mota2012,Patterson2014,Han2014,Ravazzi2015,Han2015SPARS,HanZhuNiuBaron2016ICASSP}, in which there are $P$ {\em distributed nodes} (processor nodes) and a {\em fusion center}. Each distributed node stores $\frac{M}{P}$ rows of the matrix $\A$, and acquires the corresponding linear measurements of the underlying signal $\x$. Without loss of generality, we model the measurement system in distributed node $p\in \{1,...,P\}$ as
 \begin{equation}\label{eq:one-node-meas}
    y_i=\A_i \x+z_i,\ i\in \left\{\frac{M(p-1)}{P}+1,...,\frac{Mp}{P}\right\},
 \end{equation}
 where $\A_i$ is the $i$-th row of $\A$, and $y_i$ and $z_i$ are the $i$-th entries of $\y$ and $\z$, respectively.
Once every $y_i$ is collected, we run distributed algorithms among the fusion center and $P$ distributed nodes to reconstruct the signal $\x$.
MP versions of AMP (MP-AMP) for MP-CS have been studied in the literature~\cite{Han2014,HanZhuNiuBaron2016ICASSP}.
Usually, MP platforms are designed for distributed settings such as sensor networks~\cite{pottie2000,estrin2002} or large-scale ``big data" computing systems~\cite{EC2}. We reduce the communication costs of MP platforms by applying lossy compression~\cite{Berger71,Cover06,GershoGray1993} to the communication portion of MP-AMP. 

In this paper, we consider a rich design space that includes various costs, such as the number of iterations $T$, aggregate coding rates $R_{agg}$ (defined later in~\eqref{eq:R_agg}), and the mean squared error (MSE) achieved by the reconstruction algorithm. In such a rich design space, reducing any cost is likely to incur an increase in other costs, and it is difficult to simultaneously minimize all the costs. 
Han et al.~\cite{Han2014} reduce the communication costs, and Ma et al.~\cite{MaBaronNeedell2014} develop an algorithm with reduced computation; both works~\cite{Han2014,MaBaronNeedell2014} achieve a reasonable MSE. However, the optimal trade-offs in this rich design space are not studied. 
We pose the problem of finding the best trade-offs among the individual costs $T,\ R_{agg}$, and $\text{MSE}$ as a multi-objective optimization problem (MOP), and study the properties of the Pareto optimal tuples $(T^*,R_{agg}^*,\text{MSE}^*)$~\cite{DasDennisPareto1998} of this MOP. (Note that we do not intend to provide a practical implementation to achieve the optimal trade-offs.)
Finally, we conjecture that the combined cost of computation and communication scales as $O(\log^2(1/(\text{MSE}-\text{MMSE}))$; these properties are verified numerically using
a dynamic programming (DP, cf. Bertsekas~\cite{bertsekas1995}) scheme from our prior work~\cite{ZhuBaronMPAMP2016ArXiv}.

\section{Background}\label{sec:setting}
\subsection{Centralized CS using AMP}
In the linear system~\eqref{eq:matrix_channel}, we consider an independent and identically distributed (i.i.d.) Gaussian measurement matrix $\A$, i.e., $\A_{i,j}\sim\mathcal{N}(0,\frac{1}{M})$. The signal entries follow an i.i.d. {\em Bernoulli Gaussian} distribution,
\begin{equation}
x_j\sim \epsilon \mathcal{N}(0,1)+(1-\epsilon)\delta(x_j),\label{eq:BG}
\end{equation}
where $\delta(\cdot)$ is the Dirac delta function\, and $\epsilon$ is called the {\em sparsity rate} of the signal. The noise entries obey $z_i\sim\mathcal{N}(0,\sigma_Z^2)$, where $\sigma_Z^2$ is the noise variance. Note that the results in this paper can be easily extended to priors other than~\eqref{eq:BG}.

Starting from ${\bf x}_0={\bf 0}$, the AMP framework~\cite{DMM2009} proceeds iteratively according to
\begin{align}
{\bf x}_{t+1}&=\eta_t({\bf A}^{\mathcal{T}}{\bf r}_t+{\bf x}_t)\label{eq:AMPiter1},\\
{\bf r}_t&={\bf y}-{\bf Ax}_t+\frac{1}{\kappa}{\bf r}_{t-1}
\langle d\eta_{t-1}({\bf A}^{\mathcal{T}}{\bf r}_{t-1}+{\bf x}_{t-1})\rangle\label{eq:AMPiter2},
\end{align}
where $\eta_t(\cdot)$ is a denoising function, $d\eta_{t}(\cdot)=\frac{d \eta_t({\cdot})}{d\{\cdot\}}$ is shorthand for the derivative of $\eta_t(\cdot)$, and~$\langle{\bf u}\rangle=\frac{1}{N}\sum_{i=1}^N u_i$
for some vector~${\bf u}\in\mathbb{R}^N$. The subscript $t$ represents the iteration index, $\mathcal{T}$ denotes transpose, and $\kappa=\frac{M}{N}$ is the measurement rate.
Owing to the decoupling effect~\cite{Tanaka2002,GuoVerdu2005,GuoWang2008}, in each AMP iteration~\cite{Bayati2011,Montanari2012},
the vector~$\f_t={\bf A}^{\mathcal{T}}{\bf r}_t+{\bf x}_t$
in (\ref{eq:AMPiter1}) is statistically equivalent to
the input signal ${\bf x}$ corrupted by AWGN ${\bf w}_t$ generated by a source $W\sim \mathcal{N}(0,\sigma_t^2)$,
\begin{equation}\label{eq:equivalent_scalar_channel}
\f_t=\x+{\bf w}_t.
\end{equation}
In large systems ($N\rightarrow\infty, \frac{M}{N}\rightarrow \kappa$), a useful property of AMP~\cite{Bayati2011,Montanari2012} is that
the noise variance $\sigma_t^2$ evolves following state evolution (SE):
$\sigma_{t+1}^2=\sigma^2_Z+\frac{1}{\kappa}\text{MSE}(\eta_t,\sigma_t^2)$, where the mean squared error
$\text{MSE}(\eta_t,\sigma_t^2)=\mathbb{E}_{X,W}\left[\left( \eta_t\left( X+W \right)-X \right)^2\right]$, $\mathbb{E}_{X,W}(\cdot)$ is expectation with respect to $X$ and $W$, and $X\sim f_X$ is the source that generates $\x$. Note that $\sigma_1^2=\sigma_Z^2+\frac{\mathbb{E}[X^2]}{\kappa}$, because of the all-zero initial estimate for $\x$.
Formal statements for SE appear
in prior work~\cite{Bayati2011,Montanari2012}.

This paper considers the Bayesian setting, in which we assume knowledge of the true prior for the signal $\x$. Therefore, the MMSE-achieving denoiser is the conditional expectation, $\eta_t(\cdot)=\mathbb{E}[\x|\f_t]$, which can be easily obtained.
Other denoisers such as soft thresholding~\cite{DMM2009,Montanari2012,Bayati2011} yield MSE's that are greater than that of the Bayesian denoiser.
When the true prior for $\x$ is unavailable, parameter estimation techniques
can be used.

\subsection{MP-CS using lossy MP-AMP}\label{sec:MP-CS_for_MP-AMP}
In the sensing problem formulated in~\eqref{eq:one-node-meas}, the measurement matrix is stored in a distributed manner in each distributed node. Lossy MP-AMP~\cite{HanZhuNiuBaron2016ICASSP,ZhuBaronMPAMP2016ArXiv} iteratively solves MP-CS problems using lossily compressed messages:
\begin{equation}
\mbox{Distributed nodes:}\ {\bf r}_t^p={\bf y}^p-\A^p\x_t+\frac{1}{\kappa}{\bf r}_{t-1}^p
g_{t-1},\label{eq:slave1}
\end{equation}
\begin{equation}
\quad \quad \quad {\bf f}_t^p=\frac{1}{P}\x_t+(\A^p)^{\mathcal{T}}{\bf r}_t^p,\label{eq:slave2}
\end{equation}
\begin{equation}
\mbox{Fusion center:}\ {\bf f}_{Q,t}=\sum_{p=1}^P Q({\bf f}_{t}^p),\ g_{t}=\langle d\eta_{t}({\bf f}_{Q,t})\rangle,\label{eq:master0}
\end{equation}
\begin{equation}
 \x_{t+1}=\eta_{t}( {\bf f}_{Q,t}),\label{eq:master}
\end{equation}
where $Q(\cdot)$ denotes quantization, and 
an MP-AMP iteration refers to the process from~\eqref{eq:slave1} to~\eqref{eq:master}. 
The reader might notice that the fusion center also needs to transmit the denoised signal vector $\x_t$ and a scalar $g_{t-1}$ to the distributed nodes. The transmission of the scalar $g_{t-1}$ is negligible, and the fusion center may broadcast $\x_t$ so that naive compression of $\x_t$, such as compression with a fixed quantizer, is sufficient. Hence, we will not discuss possible lossy compression of the messages transmitted by the fusion center. 

Assume that we quantize $\f_t^p, \forall p$, and use $C$ bits to encode the quantized vector $Q(\f_t^p)\in\mathbb{R}^N$. The {\em coding rate} is $R=\frac{C}{N}$. We incur a {\em distortion} (or quantization error) $D_t=\frac{1}{N}\sum_{i=1}^N(Q(f_{t,i}^p)-f_{t,i}^p)^2$ at iteration $t$ in each distributed node,\footnote{Because we assume that the matrix $\A$ and noise $\z$ are both i.i.d., the expected distortions are the same over all $P$ nodes, and can be denoted by $D_t$ for simplicity. Other distortion metrics $d(\cdot,\cdot)$ can also be used~\cite{Cover06}.} where $Q(f_{t,i}^p)$ and $f_{t,i}^p$ are the $i$-th entries of the vectors $Q(\f_t^p)$ and $\f_t^p$, respectively.
The rate distortion function, denoted by $R(D)$, offers the fundamental information theoretic limit on the coding rate $R$ for communicating a sequence up to distortion $D$~\cite{Cover06,Berger71,GershoGray1993,WeidmannVetterli2012}.
A pivotal conclusion from RD theory is that coding rates can be greatly reduced even if $D$ is quite small.
The function $R(D)$ can be computed in various ways~\cite{Arimoto72,Blahut72,Rose94}, and can be achieved by an RD-optimal quantization scheme. Other quantization schemes require larger coding rates to achieve the same expected distortion $D$.

The goal of this paper is to understand the fundamental trade-offs for MP-CS using MP-AMP. Hence,
throughout this paper, we assume that
appropriate vector quantization (VQ) schemes~\cite{LBG1980,Gray1984,GershoGray1993} that achieve $R(D)$ are applied within each MP-AMP iteration, although our analysis is readily extended to practical quantizers such as scalar quantizer with entropy coding~\cite{GershoGray1993,Cover06}.
Therefore, the signal {\em at the fusion center} before denoising can be modeled as
\begin{align}
\f_{Q,t}=\sum_{p=1}^P Q(\f_t^p)=\x+{\bf w}_t+{\bf n}_t,\label{eq:indpt_noises}
\end{align}
where ${\bf w}_t$ is the equivalent scalar channel noise~\eqref{eq:equivalent_scalar_channel} and ${\bf n}_t$ is the overall quantization error whose entries follow $\mathcal{N}(0,PD_t)$. 
For large block sizes, we expect the VQ quantization error, ${\bf n}_t$, to resemble Gaussian noise, which is independent of $\x+{\bf w}_t$.
The SE for the lossy MP-AMP~\cite{ZhuBaronMPAMP2016ArXiv,HanZhuNiuBaron2016ICASSP} follows
\begin{equation}
\sigma_{t+1}^2=\sigma^2_Z+\frac{1}{\kappa}\text{MSE}(\eta_t,\sigma_t^2+PD_t),\label{eq:SE_Q}
\end{equation}
where $\sigma_t^2$ can be estimated by
$\widehat{\sigma}_t^2 = \frac{1}{M}\|{\bf r}_t\|_2^2$ with $\|\cdot\|_p$ denoting the $\ell_p$ norm~\cite{Bayati2011,Montanari2012}, and $\sigma_{t+1}^2$ is the variance of ${\bf w}_{t+1}$.
The rigorous justification of~\eqref{eq:SE_Q} by extending Bayati and Montanari~\cite{Bayati2011} is left for future work.

Denote the coding rate used to transmit $Q(\f^p_t)$ at iteration $t$ by $R_t$. The sequence of $R_t,\ t=1,...,T$, where $T$ is the total number of MP-AMP iterations, is called
the {\em coding rate sequence}, and is denoted by the vector $\mathbf{R}=[R_1,...,R_T]$. Given the coding rate sequence $\mathbf{R}$, the distortion $D_t$ can be evaluated with $R(D)$, and the scalar channel noise variance $\sigma_t^2$ can be evaluated with~\eqref{eq:SE_Q}.
Hence, the MSE for $\mathbf{R}$ can be predicted; we call it SE-predicted MSE. The MSE at the last iteration is called the {\em final MSE}.

\section{Achievable Performance Region}\label{sec:Pareto}
Following the discussion of Sec.~\ref{sec:setting}, we can see that the lossy compression of ${\bf f}_t^p, \forall p \in \{1,...,P\}$, can reduce communication costs. On the other hand, the greater the savings in the coding rate sequence $\mathbf{R}$, the worse the final MSE is expected to be. If a certain level of final MSE is desired under a small budget of coding rates, more iterations $T$ will be needed. Define the aggregate coding rate $R_{agg}$ as the sum of all the coding rates in $\mathbf{R}$,
\begin{equation}\label{eq:R_agg}
R_{agg}=\| \mathbf{R} \|_1=\sum_{t=1}^T R_t.
\end{equation}
As mentioned above, there is a trade-off between $T$, $R_{agg}$, and the final MSE, and there is no  optimal solution that minimizes them simultaneously.
To deal with such trade-offs in a multi-objective optimization (MOP) problem, it is customary to think about the concept of {\em Pareto optimality}~\cite{DasDennisPareto1998}.

\subsection{Properties of achievable region}\label{sec:property}
Define the computation cost rate, $C_1$, as the cost of computation in one MP-AMP iteration, and define the communication cost rate, $C_2$, as the cost of transmitting 1 bit for $Q({\bf f}_t^p)$~\eqref{eq:master0}. We further define the {\em relative cost} as 
\begin{equation}\label{eq:relativeCost}
b=\frac{C_1}{C_2}.
\end{equation}
For notational convenience, denote by ${\cal E}(T, R_{agg})$ all of the MSE values that would be provided by the pair $(T,R_{agg})$ for some relative cost $b$~\eqref{eq:relativeCost}, among which the smallest MSE is denoted by $\text{MSE}^*(R_{agg},T)$.
Furthermore, 
the achievable set
$\cal C$ is defined as\footnote{$\mathbb{R}_{\geq 0}$ denotes the set of non-negative real numbers.}
$$
{\cal C} := \{(T,R_{agg}, \text{MSE}) \in R_{\geq 0}^3: \text{MSE} \in {\cal E}(T, R_{agg})\},
$$
i.e., there exists an instantiation of the MP-AMP algorithm that could reconstruct the signal with $T$ iterations and an aggregate coding rate $R_{agg}$, and yield a certain MSE.
%denote the set of all tuples $(T,R_{agg},E)$ by the {\em achievable set},
%\begin{equation}\label{eq:setC}
%\mathcal{C}=\{(T,R_{agg},E): (T, R_{agg},E) \in \mathbb{R}_{\geq 0}^3\},
%\end{equation}

\begin{mydef}\label{def:Pareto}
{\em  The point $\mathcal{X}_1\in\mathcal{C}$ is said to dominate another point $\mathcal{X}_2\in\mathcal{C}$, denoted by $\mathcal{X}_1\prec \mathcal{X}_2$, if and only if $T_1\leq T_2$, $R_{agg_1}\leq R_{agg_2}$, and $\text{MSE}_1\leq \text{MSE}_2$. A point $\mathcal{X}^*\in \mathcal{C}$ is said to be Pareto optimal if and only if there does not exist $\mathcal{X}\in \mathcal{C}$ satisfying $\mathcal{X}\prec \mathcal{X}^*$.} Furthermore, let $\mathcal{P}$ denote the set of all Pareto optimal points,
\begin{equation}\label{eq:setP}
\mathcal{P} := \{\mathcal{X}\in \mathcal{C}: \text{$\mathcal{X}$ is Pareto optimal}\}.
\end{equation}
\end{mydef}
In words, the tuple $(T,R_{agg},\text{MSE})$ is Pareto optimal if no other tuple $(T',R'_{agg},\text{MSE}')$ exists such that $T'\leq T$, $R'_{agg}\leq R_{agg}$, and $\text{MSE}'\leq \text{MSE}$. 
These are the tuples that belong to the boundary of $\cal C$.
%According to Definition~\ref{def:Pareto}, we know that the points in $\mathcal{P}$ consist of the surface of set $\mathcal{C}$.

We extend the definition of the number of iterations $T$ to a probabilistic one. We assume that the number of iterations is drawn from a probability distribution $\pi$ over $\mathbb{N}$, such that $\sum_{i=1}^{\infty} \pi_i = 1$. Of course, this definition contains a deterministic $T = j$ as a special case with $\pi_j = 1$ and $\pi_i =0$ for all $i \neq j$.
%We further define the function $\red{\underline{R}(\cdot,\cdot): \left(\mathbb{R}_{\geq 0}\right)^2 \to \mathbb{R}_{\geq 0}}$ as the Pareto optimal function implicitly as \red{
%$\underline{R}(T,E) = {R}_{agg}  \Leftrightarrow (T,R_{agg},E) \in \mathcal{P}.$}
Armed with this definition of Pareto optimality and the probabilistic definition of the number of iterations, we have the following lemma.

\begin{mylemma}\label{th:convex1}
{\it For a fixed noise variance $\sigma^2_Z$, measurement rate $\kappa$, and $P$ distributed nodes in MP-AMP, the achievable set $C$ is a convex set.}
\end{mylemma}
\begin{IEEEproof}
We need to show that for any $(T^{(1)},R^{(1)}_{agg},\text{MSE}^{(1)})$, $(T^{(2)},R^{(2)}_{agg},\text{MSE}^{(2)})$ $\in \mathcal{C}$ and any $0<\lambda<1$,
\begin{equation}\label{eq:suff1}
\begin{split}
(\lambda T^{(1)} + &(1-\lambda)T^{(2)},\lambda R^{(1)}_{agg} + (1-\lambda)R^{(2)}_{agg},\\
&\lambda \text{MSE}^{(1)}+(1-\lambda)\text{MSE}^{(2)}) \in \mathcal{C}.
\end{split}
\end{equation}

\begin{figure}[t]
\begin{center}
\includegraphics[width=8cm]{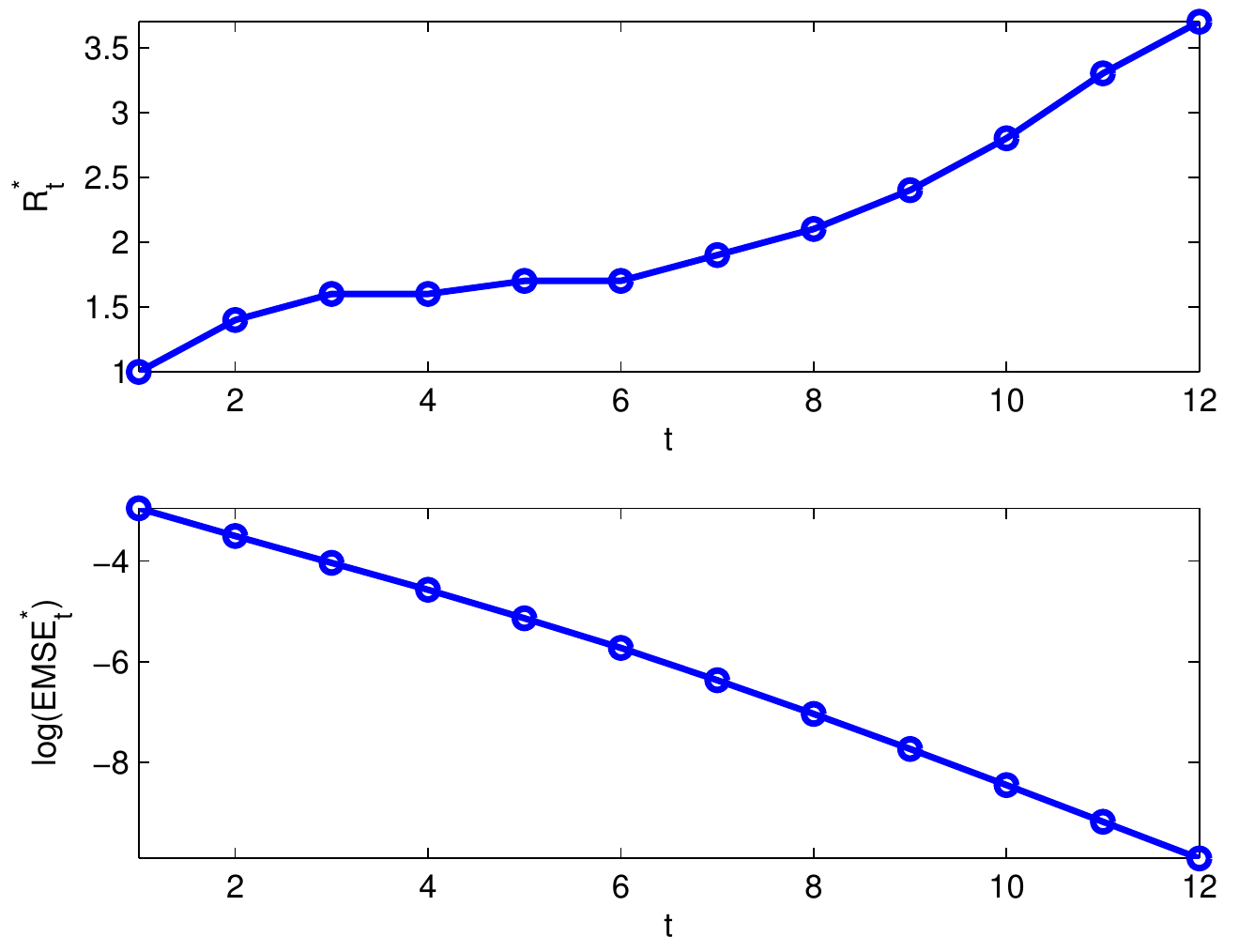}
\end{center}
\caption{The optimal coding rate sequence $\mathbf{R}$ (top panel) and $\text{EMSE}^*_t$ (bottom) are  shown as functions of $t$. ($\kappa=0.4$, $P=100$, $\sigma_Z^2=\frac{1}{400}$, and $b=2$.)}
\label{fig:RtAndEMSE}
\end{figure}

We use the well-known time-sharing argument (see Cover and Thomas~\cite{Cover06}).
Assume that $(T^{(1)},R^{(1)}_{agg},\text{MSE}^{(1)})$, $(T^{(2)},R^{(2)}_{agg},\text{MSE}^{(2)}) \in \mathcal{C}$ are achieved by probability distributions $\pi^{(1)}$ and $\pi^{(2)}$, respectively.
Let us select all the parameters of the first tuple with probability $\lambda$ and those of the second tuple with probability $(1-\lambda)$.
Hence, we have
 $\pi = \lambda \pi^{(1)} + (1-\lambda) \pi^{(2)}$. Due to the linearity of expectation, we have $T = \lambda T^{(1)} + (1-\lambda) T^{(2)}$, and $\text{MSE} = \lambda \text{MSE}^{(1)} + (1-\lambda) \text{MSE}^{(2)}$. Again, due to the linearity  of expectation, $R_{agg} = \lambda R^{(1)}_{agg} + (1-\lambda)R^{(2)}_{agg}$, implying that~\eqref{eq:suff1} is satisfied, and the proof is complete.
\end{IEEEproof}

\begin{mydef}\label{def:funcs}
{\em Let the function $R^*(T,\text{MSE}):\mathbb{R}_{\geq 0}^2\rightarrow \mathbb{R}_{\geq 0}$ be the Pareto optimal rate function, which is implicitly described as $R^*(T,\text{MSE})=R_{agg}^* \Leftrightarrow (T,R_{agg}^*,\text{MSE}) \in \mathcal{P}$. We further define implicit functions $T^*(\text{MSE},R_{agg})$ and $\text{MSE}^*(R_{agg},T)$ in a similar way.}
\end{mydef}

\begin{corollary}
The functions $R^*(T,\text{MSE})$, $T^*(\text{MSE},R_{agg})$, and $\text{MSE}^*(R_{agg},T)$ are convex in their arguments.
\end{corollary}

Note that our proof for the convexity of the set $\mathcal{C}$ might be extended to other distributed iterative learning algorithms that might use lossy compression.

\begin{figure*}[t]
  \subfigure[]{
    \label{fig:rateVSiter} %% label for first subfigure
    \includegraphics[width=6.2cm]{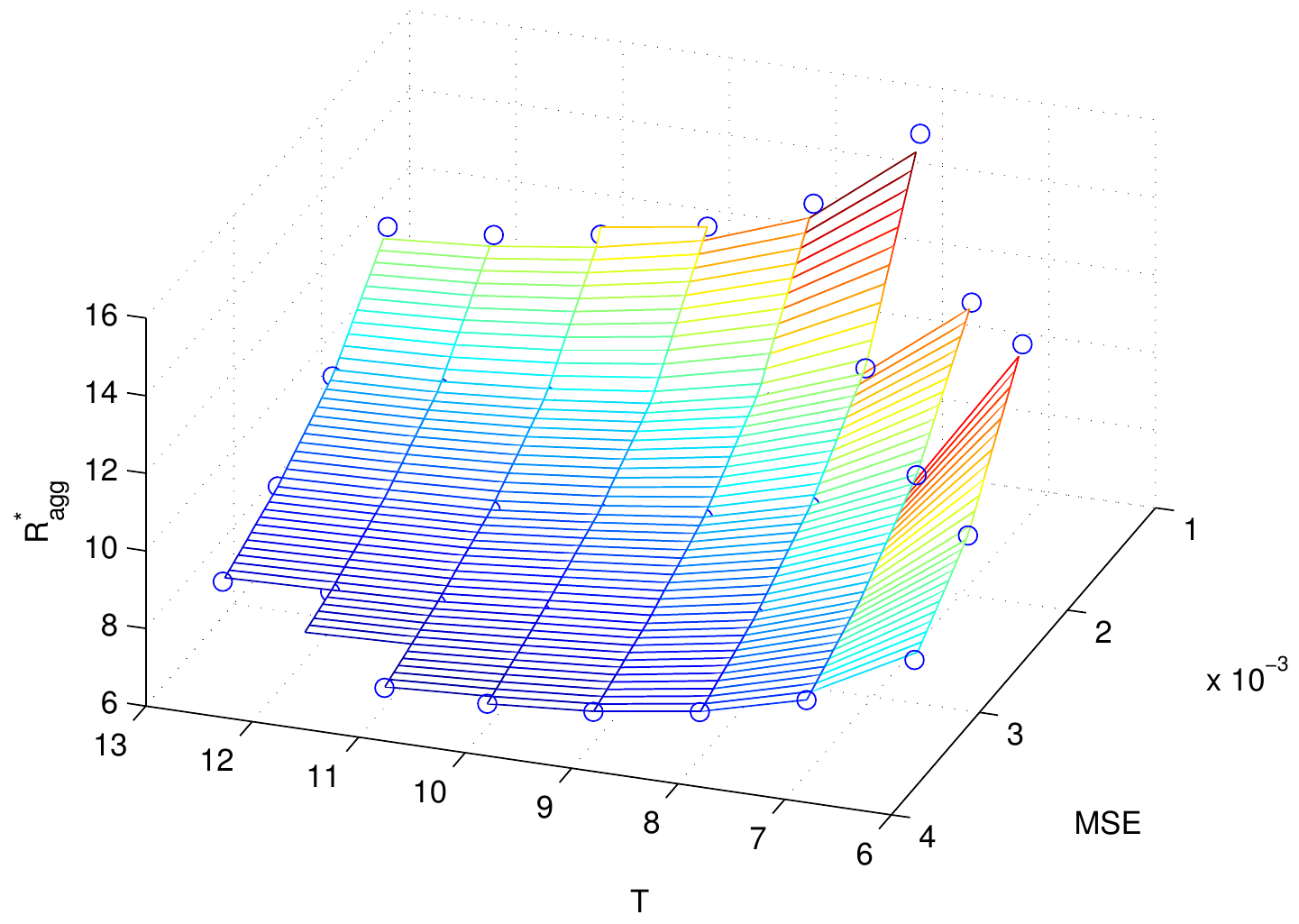}}  
  \subfigure[]{
    \label{fig:Pareto2d_fixT} %% label for second subfigure
    \includegraphics[width=5.6cm]{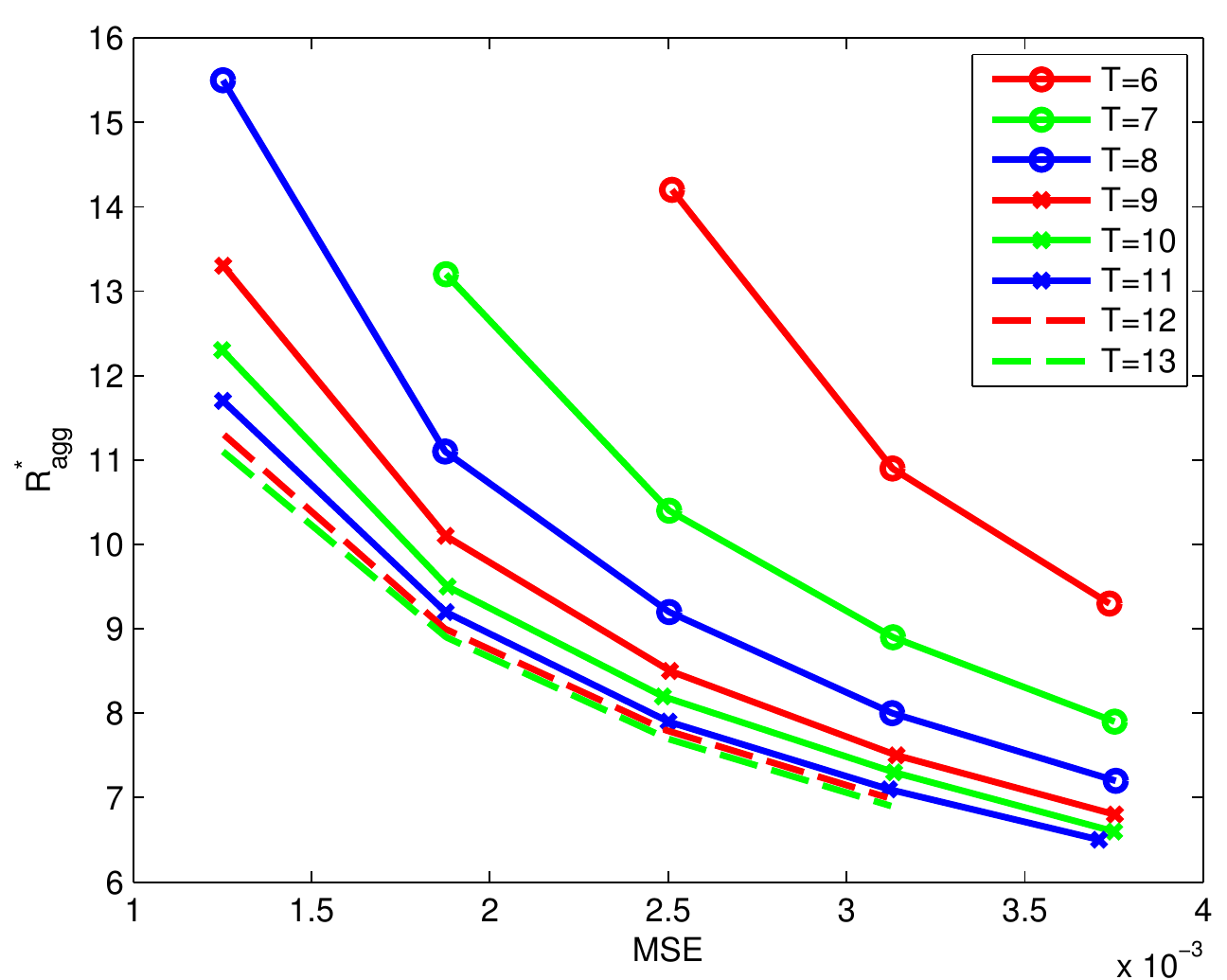}}
  \subfigure[]{
    \label{fig:Pareto2d_fixMSE} %% label for second subfigure
    \includegraphics[width=5.6cm]{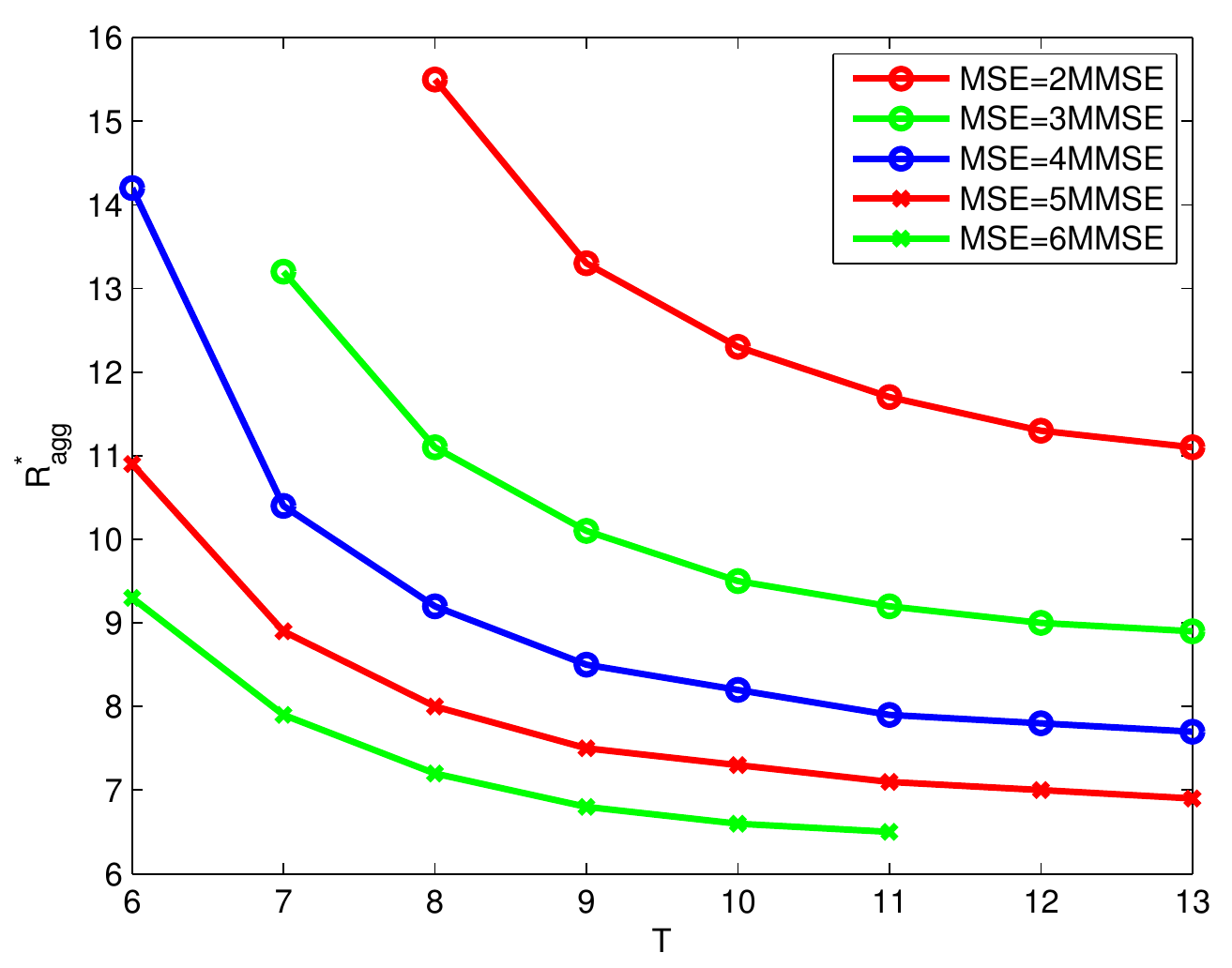}}
\caption{Pareto optimal results provided by unconstrained DP under a variety of relative costs~\eqref{eq:relativeCost}: (a) Pareto optimal surface, (b) Pareto optimal aggregate coding rate $R_{agg}^*$~\eqref{eq:R_agg} versus the achieved MSE for different optimal MP-AMP iterations $T$, and (c) Pareto optimal $R_{agg}^*$~\eqref{eq:R_agg} versus the number of iterations $T$ for different optimal MSE. The signal is Bernoulli Gaussian~\eqref{eq:BG} with $\epsilon=0.1$. ($\kappa=0.4$, $P=100$, and $\sigma_Z^2=\frac{1}{400}$.)}\label{fig:Pareto}
\end{figure*}

\subsection{Limiting performance of MP-AMP}
These discussions raise the question whether we can provide some asymptotic analysis of the achievable region. We believe that such an analysis is indeed possible in the limit of MSE that approaches the MMSE. 
Define the excess MSE ($\text{EMSE}$)~\cite{MaBaronBeirami2015ISIT},
$\text{EMSE}=\text{MSE}-\text{MMSE}$.
Consider a case where we aim to reach a very low $\text{EMSE}$. Montanari~\cite{Montanari2012} provided a graphical interpretation of the relation between the MSE performance of AMP in iteration $t$ and the statistical properties of the denoiser $\eta_t(\cdot)$ being used.
In the limit of small $\text{EMSE}$, the $\text{EMSE}$ decreases by a nearly-constant multiplicative factor in each AMP iteration, yielding a geometric decay of the MMSE.
In MP-AMP, in addition to the equivalent scalar channel noise ${\bf w}_t$, we have additive quantization error ${\bf n}_t$~\eqref{eq:indpt_noises}.
In order for the $\text{EMSE}$ in an MP-AMP system to decay geometrically, the quantization error $D_t$ must decay at least as quickly. To obtain this geometric decay in $D_t$, 
recall that in the high resolution limit, the distortion-rate function typically takes the form $D(R)=C_32^{-2R}$~\cite{GershoGray1993}, where $C_3>0$ is some constant.
We propose for $R_t$ to have the form, 
\begin{equation}
\label{eq:speculated_Rt}
R_t=C_4+C_5t, 
\end{equation}
where $C_4$ and $C_5$ are constants. This rate will not yield a distortion that decays exactly geometrically, because the distribution of $\f_t^p$
will be dependent on $t$. That said, in the limit of small $\text{EMSE}$, the distribution barely changes between iterations, and so it is plausible to expect 
$D_t\approx C_62^{-C_7t}[1+o_t(1)]$, 
where the decay rate $C_7$ is a function of the extra coding rate $C_5$ per iteration (\ref{eq:speculated_Rt}), 
and the multiplicative term $1+o_t(1)$ converges to 
1 in the limit of large $t$, because the distribution barely changes between iterations for large $t$.
Now that we have driven down the quantization error geometrically, we conjecture that the Pareto optimal EMSE, $\text{EMSE}^*\!=\!\text{MSE}^*\!-\!\text{MMSE}$, decays at the same rate,
\begin{equation}
\label{eq:speculated_EMSEt}
\text{EMSE}_t^*\approx C_82^{-C_7t}[1+o_t(1)].
\end{equation}
Combining~\eqref{eq:speculated_Rt} and~\eqref{eq:speculated_EMSEt}, and considering the definition of $R_{agg}$~\eqref{eq:R_agg}, the total computation and communication cost is $O(T)+O(R_{agg})=O(T^2)$, which is $O(\log^2(1/\text{EMSE}^*))$. We have the following conjecture.
\begin{CONJ}
The total computation and communication cost scales as $O(\log^2(1/\text{EMSE}^*))$.
\end{CONJ}

Having provided this conjecture, we back it up numerically by running our unconstrained DP scheme (Sec.~\ref{sec:unconstrainedDP})~\cite{ZhuBaronMPAMP2016ArXiv} on a problem with relatively small $\text{EMSE}^*_T=5\times10^{-5}$ in the last iteration $T$.
Consider reconstructing a Bernoulli Gaussian signal~\eqref{eq:BG} with $\epsilon=0.1$. The signal is measured in an MP platform with $P=100$ distributed nodes according to~\eqref{eq:one-node-meas}.
The measurement rate is $\kappa=\frac{M}{N}=0.4$, and the noise variance is $\sigma_Z^2=\frac{1}{400}$. The relative cost is $b=2$~\eqref{eq:relativeCost}.
Fig.~\ref{fig:RtAndEMSE} illustrates the optimal coding rate sequence $\mathbf{R}$ and $\text{EMSE}^*_t$ as functions of the iteration number $t$. It is readily seen that after the first 5--6 iterations the coding rate seems near-linear, which confirms (18); and $\text{EMSE}^*_t$ decays geometrically, as predicted by~\eqref{eq:speculated_EMSEt}.

\section{Numerical results}\label{sec:discuss}
After proving that the achievable set $\mathcal{C}$ is convex, we apply the unconstrained DP developed in Zhu and Baron~\cite{ZhuBaronMPAMP2016ArXiv} to find the Pareto optimal points for various relative costs~\eqref{eq:relativeCost}, and illustrate the convexity of the achievable set.

\subsection{Unconstrained DP}\label{sec:unconstrainedDP}
The unconstrained DP~\cite{ZhuBaronMPAMP2016ArXiv} finds a coding rate sequence $\mathbf{R}$ over the MP-AMP iterations such that the final MSE is less than $\Delta$, while achieving the minimum cost $\Psi$.
The cost $\Psi$ for a given computation cost rate $C_1$ and communication cost rate $C_2$ is a function of the number of remaining iterations $(T-t)$ and the current scalar channel noise variance $\sigma^2_t$~\eqref{eq:indpt_noises}. In the basis case, $T-t=0$, the cost is $C_1+C_2R_{T}$. After solving the basis case, we iterate back in time by decreasing $t$,
\begin{equation*}
 \Psi_{T-t}(\sigma^2_t)\! =\! 
\min_{R'}\!\left\{ C_1\! \times\! {\mathbbm{1}}_{R' \neq 0}\!+\!C_2 R'\!+\!
\Psi_{T-(t+1)}(\sigma^2_{t+1}(R'\!))\right\},
\end{equation*}
%\begin{equation*}
%\begin{split}
% &\Psi_{T-t}(\sigma^2_t) = \\
%& \min_{R'}\left\{ C_1 \times {\mathbbm{1}}_{R' \neq 0}+C_2 R'+
%\Psi_{T-(t+1)}(\sigma^2_{t+1}(R'))\right\},
%\end{split}
%\end{equation*}
where $R'$ is the coding rate used in the current
MP-AMP iteration~$t$, ${\mathbbm{1}}_{\mathcal{A}}$ is the indicator function, which is 1 if the condition $\mathcal{A}$ is met, else 0, and
$\sigma^2_{t+1}(R')$ is the
variance of the noise ${\bf w}_{t+1}$
of the scalar channel~\eqref{eq:indpt_noises} in the next MP-AMP iteration after
transmitting $\f_t^p$ at rate $R'$.
A discretized search space of $\sigma^2_t$ and $R'$ is utilized~\cite{ZhuBaronMPAMP2016ArXiv}.

The coding rates $R'$ that minimize the cost function $\Psi_{T-t}(\sigma_t^2)$ for different $t$ and $\sigma^2_t$ are stored in a table $\mathcal{R}(t,\sigma^2_t)$.
After the unconstrained DP finishes, we
obtain the coding rate sequence $\mathbf{R}$ from the table $\mathcal{R}(t,\sigma^2_t)$.

\subsection{Pareto optimal points via unconstrained DP}
According to Definition~\ref{def:Pareto}, the resulting tuple $(T,\|\mathbf{R}\|_1,\Delta)$ from the unconstrained DP in Sec.~\ref{sec:unconstrainedDP} is Pareto optimal. Hence, in this subsection, we run the unconstrained DP to obtain the Pareto optimal points for a certain distributed linear system under various relative costs~\eqref{eq:relativeCost}.

Consider the same setting as in Fig.~\ref{fig:RtAndEMSE}, except that we analyze MP platforms~\cite{pottie2000,estrin2002,EC2} with a variety of relative costs~\eqref{eq:relativeCost}. Running the unconstrained DP scheme developed in Sec.~\ref{sec:unconstrainedDP},
we obtain the optimal coding rate sequence $\mathbf{R}$ that yields the lowest combined cost while helping MP-AMP achieve an MSE that is at most $\Delta\in \{2,3,...,6\}\times \text{MMSE}$. In Fig.~\ref{fig:rateVSiter}, we draw the Pareto optimal surface obtained by running the unconstrained DP; the circles on the surface are the Pareto optimal points we analyzed. Fig.~\ref{fig:Pareto2d_fixT} plots the aggregate coding rate
as a function of different MSE with different optimal numbers of MP-AMP iterations $T$.
Fig.~\ref{fig:Pareto2d_fixMSE} plots the aggregate coding rate
as a function of different $T$ with different optimal MSE.
We can see that the surface comprised of the Pareto optimal points is indeed convex.

With stricter requirements on the final MSE (meaning smaller $\Delta$), more iterations $T$ and greater aggregate coding rates $R_{agg}$~\eqref{eq:R_agg} are needed. Optimal coding rate sequences increase the coding rate
to reduce the number of iterations when communication costs are low~\cite{ZhuBaronMPAMP2016ArXiv} (examples are commerical cloud computing systems~\cite{EC2}, multi-processor CPUs, and graphic processing units), whereas more iterations allow to reduce the coding rate when communication is costly~\cite{ZhuBaronMPAMP2016ArXiv} (for example, in sensor networks~\cite{pottie2000,estrin2002}).

%\begin{myremark}
%Examining all coding rate sequences $\mathbf{R}$ in our DP results, we notice that the coding rate is monotone non-decreasing, i.e., $R_1\leq R_2\leq \cdots\leq R_T$. This seems intuitive, because in early iterations of (MP-)AMP, the scalar channel noise ${\bf w}_t^p$ is large,
%which does not require
%transmitting high fidelity $\f_t^p$ (cf.~\eqref{eq:slave2}). Hence, low rate $R_t$ is good enough. As the iterations proceed, the scalar channel noise ${\bf w}_t^p$ is reduced, and large quantization error ${\bf n}_t$~\eqref{eq:indpt_noises} will be unfavorable to the final MSE. Hence, higher rates are needed in later iterations.
%This remark coincides with~\eqref{eq:speculated_EMSEt}.
%\end{myremark}

%---------------
\section*{Acknowledgments}
%---------------
The authors thank Puxiao Han and Ruixin Niu for numerous discussions about MP settings of CS and AMP. We also thank Yanting Ma for useful suggestions.

%---------------
\end{document}